\documentclass[twocolumn,pra,aps,showpacs,superscriptaddress,floatfix]{revtex4-1}
\usepackage[latin1]{inputenc}
\usepackage{bm}
\usepackage[usenames]{color}
\usepackage{multirow}
\usepackage{amssymb}
\usepackage{amsbsy}
\usepackage{amsmath}
\usepackage{stmaryrd}
\usepackage{graphicx}
\usepackage{epsfig}
\usepackage{placeins}
\usepackage{ulem}
\makeatletter

\newcommand{\tn}{\textnormal}

\begin{document}

\title{Finite-temperature charge transport in the one-dimensional Hubbard model}

\author{F. Jin}
\affiliation{Institute for Advanced Simulation, J\"ulich Supercomputing Centre,
Forschungszentrum J\"ulich, D-52425 J\"ulich, Germany}

\author{R. Steinigeweg}
\email{rsteinig@uos.de}
\affiliation{Department of Physics, University of Osnabr\"uck, D-49069 Osnabr\"uck, Germany}
\affiliation{Institute for Theoretical Physics, Technical University Braunschweig, D-38106 Braunschweig, Germany}

\author{F. Heidrich-Meisner}
\affiliation{Department of Physics and Arnold Sommerfeld Center for Theoretical Physics,
Ludwig-Maximilians-Universit\"at M\"unchen, D-80333 M\"unchen, Germany}

\author{K. Michielsen}
\affiliation{Institute for Advanced Simulation, J\"ulich Supercomputing Centre,
Forschungszentrum J\"ulich, D-52425 J\"ulich, Germany}
\affiliation{RWTH Aachen University, D-52056 Aachen, Germany}

\author{H. De Raedt}
\affiliation{Department of Applied Physics, Zernike Institute for Advanced Materials,
University of Groningen, NL-9747AG Groningen, The Netherlands}

\date{\today}

\begin{abstract}
We study the charge conductivity of the one-dimensional repulsive Hubbard model at finite
temperature using the method of dynamical quantum typicality, focusing at half
filling. This numerical approach allows us to obtain current autocorrelation functions
from systems with as many as $18$ sites, way beyond the range of standard exact
diagonalization. Our data clearly suggest that the charge Drude weight vanishes
with a power law as a function of system size. The low-frequency dependence of
the conductivity is consistent with a finite dc value and thus with diffusion,
despite large finite-size effects. Furthermore, we consider the mass-imbalanced
Hubbard model for which the charge Drude weight decays exponentially with system
size, as expected for a non-integrable model. We analyze the conductivity and
diffusion constant as a function of the mass imbalance and we observe that the
conductivity of the lighter component decreases exponentially fast with the
mass-imbalance ratio. While in the extreme limit of immobile heavy particles, the
Falicov-Kimball model, there is an effective Anderson-localization mechanism leading
to a vanishing conductivity of the lighter species, we resolve finite conductivities  
for an inverse mass ratio of $\eta \gtrsim 0.25$.
\end{abstract}

\maketitle

\section{Introduction}

The Hubbard model is a paradigmatic model in the theory of strongly correlated
electrons, capturing some of the essential many-body effects due to short-range
electronic correlations in condensed matter physics: Mott-insulating behavior
and the resulting localization of magnetic moments with antiferromagnetic
spin correlations. Moreover, the Hubbard model is the parent Hamiltonian for
the Heisenberg and t-J model, which describe its low-energy physics in the strongly
interacting regime \cite{dagotto94,lee06,esslinger}.

The interest in the one-dimensional (1D) version of the model arises because
of the existence of an exact solution based on the Bethe ansatz \cite{essler-book}
and its relevance for quasi-1D materials \cite{jerome,vescol,hasegawa97,claessen04,wall11},
nanostructures \cite{bockrath99,ishii03,deshpande09} and realizations
with ultracold atomic gases in optical lattices \cite{esslinger,pertot14}. A
recent optical-lattice experiment has investigated the {\it non-equilibrium}
charge transport in the two-dimensional Hubbard model \cite{schneider12}.

The Hamiltonian of the 1D repulsive Hubbard model is given by  $H=\sum_{l=1}^{L} h_l$
with local terms
\begin{equation}\label{eq:h}
h_l= -t_\text{h} \sum_{\sigma}\left(c_{l,\sigma}^\dagger c_{l+1,\sigma}^{\phantom{\dagger}}
+ \tn{h.c.} \right) + {U}(n_{l,\uparrow}-\frac{1}{2})(n_{l,\downarrow}-\frac{1}{2})
\end{equation}
with $c_{L+1,\sigma} = c_{1,\sigma}$, where $c_{l,\sigma}$ ($c_{l,\sigma}^\dagger$) annihilates (creates)
a fermion with spin $\sigma=\uparrow,\downarrow$ on site $l$, and $n_{l,\sigma}=
c_{l,\sigma}^\dagger c_{l,\sigma}^{\phantom{\dagger}}$ is the local density.
$L$ is the  number of sites, $t_\text{h}$ is the hopping matrix element, and $U$
denotes the on-site Coulomb repulsion.

Despite the success of the theory of such integrable systems in computing many
equilibrium properties, the quantitative and qualitative understanding of
transport within linear response theory  has proven to be a hard problem
\cite{zotos-review,hm07}. While the zero-temperature transport properties are
completely understood (see, e.g., \cite{kirchner99}), the main open questions
concern transport of charge, spin, or energy at finite temperatures $T > 0$.
The theory of the algebraic structure of the Bethe ansatz provides knowledge
of local conservation laws, which can give rise to ballistic transport \cite{zotos97}.

This ballistic transport is usually described via the Drude weight $D$, the
zero-frequency contribution in the real part of the conductivity $\sigma(\omega)$,
\begin{equation}\label{eq:sigma}
\mbox{Re}\, \sigma(\omega) = 2 \pi D \delta(\omega) + \sigma_{\rm reg}(\omega) \,.
\end{equation}
As was argued by Zotos, Naef, and Prelov{\v{s}}ek  \cite{zotos97}, a finite Drude
weight exists if a lower bound is obtained from the Mazur inequality 
\begin{equation}\label{eq:bound}
D_{} \geq \frac{1}{2T L}\sum_i \frac{\langle Q_i j\rangle^2}{\langle Q_i^2
\rangle}\, ,
\end{equation}
where $\langle \bullet \rangle$ is the thermodynamic average at temperature
$T$. Such a bound exists if at least one conserved charge $Q_i$ has a finite
overlap with the current operator $j$. The $Q_i=\sum_l q_{l,i}$ are commonly
ordered by their range, $i=1$ corresponding to particle number $Q_1= N$ and
$i=2$ corresponding to the Hamiltonian $Q_2 = H$. $Q_3$ has range three (i.e.,
$q_{l,3}$ involves operators acting on three neighboring sites) and has the
same structure as the energy-current operator, yet the two differ in the
prefactor of one term \cite{zotos97}. As a consequence, thermal transport in
the one-dimensional Hubbard model is ballistic at any finite temperature
$T > 0$ \cite{zotos97,karrasch15a}. Recently, it has been shown that there
are also quasi-local conserved quantities in Bethe-ansatz integrable systems which
can be crucial for some transport channels \cite{prosen11,prosen13,mierzejewski15}.
Using the Mazur inequality, one obtains a non-zero Drude weight for charge
transport for any filling $n=N/L$ ($N$ is the  number of fermions)
{\it other} than $n=1/2$, from considering only the leading non-trivial
local conserved charge $Q_3$ of range three. The case of half filling has been
discussed controversially, with some studies arguing in favor of a finite charge
Drude weight $D > 0$ \cite{fujimoto97,kirchner99} while others provided
evidence for a vanishing $D = 0$ \cite{peres00,carmelo12,karrasch14a} or at best a
very small $D$ \cite{karrasch14a} in the thermodynamic limit (tDMRG gives a small
upper bound to the Drude weight). The situation
thus appears to be similar to spin transport in the spin-1/2 XXZ chain at
zero magnetization, where also no local conservation law yields a non-zero
bound to the spin Drude weight \cite{zotos97}, while numerical results \cite{zotos96,
narozhny98,hm03,herbrych11,karrasch12,karrasch13,steinigeweg14} and Bethe-ansatz
based calculations \cite{zotos99,benz05} strongly indicate a nonzero spin Drude
weight at least in its gapless phase, with the possible exception of the point
of full SU(2) symmetric exchange, i.e., the Heisenberg chain. For that model,
though, quasi-local conservation laws have ultimately been identified as being
at the heart of the ballistic spin transport \cite{prosen11,prosen13} at zero
magnetization and in its gapless phase.

The connection between (quasi-)local conservation laws and ballistic transport
is closely related to how such conservation laws  affect thermalization in
integrable systems \cite{rigol07}. Consider a quantum quench in which the force
driving the current is turned off. If this initial condition leads to a finite
value of $\langle j Q_i \rangle$, then the current will never completely decay
back to zero. A simple example is the quench of a flux piercing a ring, which
has been studied in this context \cite{mierzejewski14}.

Besides the question of the (divergent) zero-frequency contribution, the actual
frequency dependence of the optical conductivity $\sigma_{\rm reg}(\omega)$
constitutes an equally interesting problem \cite{prelovsek04,steinigeweg11a,
steinigeweg12a,karrasch14a}. Some insight can be gained from effective low-energy
theories such as bosonization \cite{giamarchi91,sirker09,sirker11}, which is, however,
limited to very low temperatures and may not correctly capture effects due to
integrability without fine-tuning of parameters. An exact diagonalization study
observed strong anomalous finite-size effects in $\sigma_{\rm reg}(\omega)$ of
integrable Mott insulators \cite{prelovsek04}, while many studies conclude that
the dc conductivity
\begin{equation}
\sigma_{\rm dc} = \lim_{\omega\to 0} \sigma_{\rm reg}(\omega)
\end{equation}
is nonzero in such systems \cite{prelovsek04,karrasch14a}. A recent density matrix
renormalization group study suggests a generic divergence of $\sigma_{\rm dc}(T)$
at low temperatures with $\sigma_{\rm dc}\propto 1/T $ \cite{karrasch14a}, different
from the Fermi-liquid behavior $\sigma_{\rm dc}\propto 1/T^2$ that emerge in
sufficiently high dimensions \cite{uhrig95}. For the high-temperature regime, a
lower bound for the diffusion constant $\mathcal{D}$ has been derived \cite{prosen14},
reading
\begin{equation}
\mathcal{D} \geq \mbox{const.} \cdot \frac{t_\text{h}^3}{ U^2}\,. \label{lowerbound}
\end{equation}
(Note that ${\cal D} \neq D$.) While our primary interest is in the behavior
in the linear response regime, we mention that numerical simulations of
boundary-driven transport through open Hubbard chains also indicate diffusive
high-temperature transport \cite{prosen12}.

In our work, we revisit the problem of charge transport in the Hubbard chain at
half filling by employing the method of dynamical quantum typicality (DQT).
Basically, this approach uses single pure states that are constructed to yield
{\it typical} thermal behavior at finite temperature to compute the time dependence
of correlation functions. In the current context of transport, this method has
recently been applied to the calculation of the spin Drude weight in XXZ chains
\cite{steinigeweg14} and to transport in various non-integrable models
\cite{steinigeweg14a,steinigeweg15,steinigeweg15a}. Since only a pure state needs
to be propagated in the DQT method, any means of propagating the wave
function such as a forward integration or Krylov-space based  approaches can be used,
giving access to system sizes as large as $L=18$, which is comparable to what can
be reached for the ground state via Lanczos methods.

We extract the Drude weight from the long-time behavior of current autocorrelation
functions and study its finite-size dependence. We observe a power-law decay with
system size to zero, which we interpret in the framework of the eigenstate
thermalization hypothesis applied to integrable systems \cite{steinigeweg13}. Thus,
our results confirm the predictions of Ref.~\cite{peres00,carmelo12}, i.e., a vanishing
Drude weight $D = 0$ at finite temperatures. We further analyze the optical conductivity,
for which our data suggest a finite $\sigma_{\rm dc}$. Depending on how the time-dependent
data are converted to frequency, one either recovers the anomalous, system-size dependent
fluctuations discussed in \cite{prelovsek04} or one obtains a smooth, diffusive-like
low-frequency dependence.

The Hubbard model can equivalently be formulated as a spin-1/2 model defined on
a two-leg ladder: spin-up and spin-down fermions live on the two separate legs,
where the exchange is of $XY$ type along the legs, while on the rungs the Hubbard
interaction translates into an Ising interaction. This reformulation is, on the
one hand, useful for numerical implementations, and on the other hand, there are
several natural ways of breaking the integrability that emerge in this picture.
Transport in various spin Hamiltonians defined on spin ladders has in fact been
intensely investigated  \cite{alvarez02b,hm03,zotos04,jung06,znidaric13,znidaric13a,
steinigeweg14a,steinigeweg15a}.

Here, we consider the {\it mass-imbalanced} Hubbard model as an example of a
non-integrable system. The local Hamiltonian now takes the form:
\begin{equation}\label{eq:hmass}
h_l= - \sum_{\sigma=\uparrow,\downarrow} \Big [ {t_\sigma} \left(c_{l,\sigma}^\dagger
c_{l+1,\sigma}^{\phantom{\dagger}} + \tn{h.c.} \right) \Big ] + {U}(n_{l,\uparrow}-
\frac{1}{2})(n_{l,\downarrow}-\frac{1}{2}) \,,
\end{equation}
i.e., we introduce different hopping matrix elements $t_{\sigma}$, $\sigma=\uparrow,
\downarrow$, for the two fermionic species. 
We define the inverse mass ratio as 
\begin{equation}
\eta = \frac{t_{\downarrow}}{t_{\uparrow}}\,.
\end{equation}
In the limit of $\eta =0$, also known as Falicov-Kimball model, one naturally
obtains perfectly insulating behavior at any temperature due to an effective
Anderson-localization mechanism. In this case, all the local density operators
$n_{l,\downarrow}$ of the heavy species become conserved quantities, i.e.,
$\lbrack H, n_{l,\downarrow} \rbrack =0$. Thus, for a given random distribution
of immobile spin-down fermions, via the interaction term $U n_{l,\uparrow}
n_{l,\downarrow}$, one effectively obtains a diagonal disorder potential for
the light fermions with local potentials $\epsilon_l = U  n_{l,\downarrow}$ drawn
from a binary distribution $\epsilon_l = 0,U$. The translational invariance of the
original model at a given density of $ n_{\downarrow}=N_{\downarrow}/L$ is restored
by averaging over many random distributions of the heavy fermions.

We are interested in the dependence of the conductivities $\sigma_{\uparrow}(\omega)$
and $\sigma_{\downarrow}(\omega)$ of the heavy and light species, respectively, as
a function of the inverse mass ratio $\eta$. First, we compute the associated Drude
weights, which vanish approximately exponentially fast with system size, as expected for
a non-integrable model \cite{hm04,beugeling14,steinigeweg13}. For intermediate values of
$\eta$, we observe a regular form of $\sigma_{\uparrow}(\omega)$ and
$\sigma_{\downarrow}(\omega)$. The dc conductivity of the heavy component appears to
simply vanish quadratically with $t_{\downarrow}$, while the presence of the
heavy fermions leads to an approximately exponential decay of the dc conductivity of
the light fermions as a function of decreasing $\eta$, which we are able to resolve
for $\eta \gtrsim 0.25$.

The mass-imbalanced Hubbard model has recently attracted renewed interest in the
context of many-body localization \cite{altman14,nandkishore15} since several
authors have considered the possibility of many-body localization in translationally
invariant systems \cite{schiulaz14,grover14,deroeck14}. In our model, interactions
could thus potentially lead to a non-trivial effect in the strongly mass-imbalanced
regime. Recent work has suggested, though, that there likely is no mass-imbalance
driven localization-delocalization transition in our model at a nonzero $\eta$, but
a quasi many-body localized behavior with anomalous diffusion at small values of
$\eta$ \cite{yao15}. These results are based on exact diagonalization with $L\leq 10$.
 Our results suggest a finite, albeit exponentially small dc conductivity at
least for $\eta \gtrsim 0.25$.

The plan of the paper is the following. Section~\ref{sec:def} summarizes the
definitions and expressions of the conductivity, the Drude weight, and current
autocorrelation functions. In Section~\ref{sec:dyntyp}, we provide a brief
introduction to the DQT method and its application to the
calculation of finite-temperature current autocorrelation functions.
Section~\ref{sec:hubbard} contains our results for the integrable Hubbard chain
at half filling, while we present our data and the discussion of the
mass-imbalanced model in Sec.~\ref{sec:mass}. We conclude with a summary and
an outlook in Sec.~\ref{sec:sum}.

\section{Definitions}
\label{sec:def}

Using the Jordan-Wigner transformation, the mass-imbalanced Fermi-Hubbard
model can equivalently be formulated as a spin-$1/2$ model defined on
a two-leg ladder,
\begin{eqnarray}
h_l = \sum_{\sigma=\uparrow,\downarrow} - 2 t_\sigma ( S_{l,\sigma}^x
S_{l+1,\sigma}^x + S_{l,\sigma}^y S_{l+1,\sigma}^y ) + U S_{l,\uparrow}^z
S_{l,\downarrow}^z \, ,
\end{eqnarray}
where spin-up and spin-down fermions live on the two separate legs and the
Hubbard interaction translates into an Ising interaction. Our numerical
implementation is formulated in the spin language.

We derive the charge current from the continuity equation \cite{zotos97},
leading to $j = j_\uparrow + j_\downarrow$ and $j_\sigma = i \sum_l [n_{l,
\sigma}, h_l]$ in the Hubbard notation. In the spin notation,
\begin{equation}
j_\sigma = -2 t_\sigma \sum_l ( S_{l,\sigma}^x S_{l+1,\sigma}^y - S_{l,\sigma}^y
S_{l+1,\sigma}^x)
\end{equation}
is the spin current in the first ($\sigma=\uparrow$) or second ($\sigma=
\downarrow$) leg. We correspondingly study the two current autocorrelation
functions at inverse temperature $\beta = 1/T$
\begin{equation}
C_\sigma(t) = \frac{\text{Re} \, \langle j_\sigma(t) j_\sigma
\rangle}{L Z} = \frac{\text{Re} \, \text{Tr}\{e^{-\beta H}
j_\sigma(t) j_\sigma\}}{L \text{Tr}\{ e^{-\beta H}\}} \, , \label{ACF}
\end{equation}
where the time argument of $j_\sigma(t)$ refers to the Heisenberg
picture, $j_\sigma(0) = j_\sigma$, and $C_\sigma(0) = t_\sigma^2/2$ in the
high-temperature limit $\beta \to 0$.

From the time dependence of $C_\sigma(t)$ we determine the quantities
\begin{equation}
\bar{C}_\sigma(t_1,t_2) = \frac{1}{t_2-t_1} \int_{t_1}^{t_2} \! \text{d}t
\, C_\sigma(t) 
\end{equation}
in a time interval $[t_1,t_2]$ where $C_\sigma(t)$ has decayed to its
long-time value $C(t_1 < t < t_2) \approx C(t\to\infty)$ and is practically
constant. Thus, the quantities $\bar{C}_\sigma(t_1,t_2)$ approximate the
finite-size Drude weights of the two legs given by
\begin{equation}
D_\sigma = \frac{1}{2 \pi} \lim_{t_2\to\infty} \bar{C}_\sigma(0,t_2) \, .
\end{equation}
We determine the frequency-dependent optical conductivity $\text{Re} \,
\sigma_{\sigma, t_{\rm max}}(\omega)$ via the finite-time Fourier transformation
\begin{equation}
\text{Re} \, \sigma_{\sigma, t_{\rm max}} (\omega) = \frac{1-
e^{-\beta \omega}}{\omega}
\int_0^{t_{\rm max}} \! \text{d}t \, e^{\imath \omega t} \, C_\sigma(t) \, .
\end{equation}
Here, the choice of a particular $t_{\rm max}$ implies a frequency resolution
$\delta \omega \approx \pi/t_{\rm max}$. In the thermodynamic limit $L \to
\infty$, $\text{Re} \, \sigma_{\sigma,t_\text{max}}(\omega)$ is a
smooth function on an arbitrarily small scale $\delta \omega \to 0$ and does
not depend on the actual value of  $t_{\rm max}$ chosen, as long as it is
large compared to the current relaxation time \cite{steinigeweg15}.
For any finite $L$, however, it is important to find a reasonable $t_\text{max}$
where finite-size effects are well controlled. In particular, for integrable
systems, finding such a $t_{\rm max}$ can be a subtle issue, as discussed
later in detail. Note that, to leading order in $\beta$, $\text{Re} \,
\sigma_\sigma(\omega) \propto \beta$ and that $\text{Re} \, \sigma_\sigma(\omega)
= \sigma_\sigma(\omega)$ in the high-temperature limit. 

If we find a $(t_{\rm max},L$) region with no significant dependence on
$t_{\rm max}$ and $L$, we extract the dc conductivity $\sigma_{\sigma,
\text{dc}}$ as the low-frequency limit
\begin{equation}
\sigma_{\sigma,\text{dc}} = \lim_{\omega \to 0} \sigma_{\sigma,
t_{\rm max}}(\omega) \, .
\end{equation}
In case of vanishing Drude weights, $\sigma_{\sigma,\text{dc}} / \chi$ is
identical to the time-dependent diffusion constant
\begin{equation}
{\cal D}_{\sigma}(t_{\rm max}) = \frac{\beta}{\chi} \int_0^{t_{\rm max}}
\! \text{d} t \, C_\sigma(t) \label{eq:diff}
\end{equation}
with $\chi$ being the static susceptibility and reading, at $\beta \to 0$,
\begin{equation}
\frac{\chi}{\beta} = \frac{\text{Tr} \{ (\sum_l S_{l,\sigma}^z)^2 \} - (\text{Tr}
\{ \sum_l S_{l,\sigma}^z \})^2}{L} = \frac{1}{4}.
\end{equation}
In the case of significant
finite-size Drude weights, however, ${\cal D}_{\sigma}(t_{\rm max})$ may not
depend on $t_{\rm max}$ and $L$, while $\sigma_{\sigma,\text{dc}}$ clearly does.
Therefore, in such cases, the time-dependent diffusion constant provides
a useful alternative for extracting transport coefficients on the basis of
finite systems. Beyond technical aspects, ${\cal D}(t)$ also has a
clear physical interpretation: It directly yields information on how spatial
variances of density profiles evolve in time \cite{steinigeweg09,langer09,
langer11,karrasch14} for any finite $L$.

\section{Dynamical Quantum Typicality}
\label{sec:dyntyp}

\subsection{Concept}

In this section we first introduce a very accurate approximation of
current autocorrelation functions. This approximation then provides
the basis for the numerical technique used throughout our work. The
central idea is to replace the trace operation $\text{Tr}\{\bullet\}
= \sum_i \langle i | \bullet | i \rangle$ in Eq.\ (\ref{ACF}) by a
single scalar product $\langle \psi | \bullet | \psi \rangle$, where
$|\psi \rangle$ is a single pure state drawn at random. Since we aim
at describing the current dynamics in the full Hilbert space,
$|\psi \rangle$ is drawn at random in the full basis. Conveniently,
$|\psi \rangle$ is randomly chosen in the eigenbasis of the 
particle number,
\begin{equation}
| \psi \rangle = \sum_N | \psi_N \rangle \, , \quad
|\psi_N \rangle = \sum_s^{d_{N}} (a_s + \imath \, b_s) \, | s
\rangle \, ,
\end{equation}
where $s = s(N)$ is a label for the eigenstates with particle
number $N$. The coefficients $a_s$ and $b_s$ are random real
numbers. To be precise, these coefficients are chosen according to a
Gaussian distribution with zero mean. Thus, the pure state $|\psi
\rangle$ is chosen according to the unitary invariant Haar measure
\cite{bartsch2009,bartsch2011} and, according to typicality
\cite{gemmer2003, goldstein2006, reimann2007, popescu2006, white2009,
sugiura2012}, a representative of the statistical ensemble.

The pure state $| \psi \rangle$, and each $| \psi_N \rangle$, correspond
to the limit of high temperatures $\beta \to 0$. We incorporate finite
temperatures $\beta \neq 0$ by introducing $| \psi_N(\beta) \rangle =
\exp(-\beta H/2) \, | \psi_N \rangle$ and rewriting the current
autocorrelation function in Eq.\ (\ref{ACF}) in the form \cite{hams2000,
bartsch2009, bartsch2011, elsayed2013, steinigeweg14, steinigeweg14a}
(skipping the index $\sigma$ for clarity)
\begin{eqnarray}
C(t) &=& \frac{\text{Re} \, \sum_{N} \langle \psi_{N}(\beta)
| j(t) \, j | \psi_N(\beta) \rangle}{L \, \sum_N \langle \psi_N(\beta)
| \psi_N(\beta) \rangle} \nonumber \\
&+& \epsilon(| \psi \rangle) \, , \label{approximation1}
\end{eqnarray}
where $\epsilon(| \psi \rangle)$ is a statistical error resulting from the
random choice of $| \psi \rangle $. This error vanishes when sampling
over several $| \psi \rangle$ is performed, i.e., $\bar{\epsilon} = 0$.

However, the central advantage of Eq.\ (\ref{approximation1}) is not the
vanishing mean error $\bar{\epsilon}  = 0$ but the knowledge about the standard
deviation of errors $\Sigma(\epsilon)$. This standard deviation is
bounded from above by \cite{bartsch2009, bartsch2011, elsayed2013,
steinigeweg14},
\begin{equation}
\Sigma(\epsilon) \leq {\cal O} \left ( \frac{\sqrt{\text{Re} \, \langle
j(t) \, j \, j(t) \, j \rangle}} {L \, \sqrt{d_\text{eff}}}
\right ) \, , \label{error}
\end{equation}
where $d_\text{eff}$ is the effective dimension of the Hilbert space.
In the limit of high temperatures $\beta \to 0$, $d_\text{eff} =
4^L$ is the full Hilbert-space dimension. Consequently, if the length
$L$ is increased, $\Sigma(\epsilon)$ decreases exponentially
fast with $L$. At arbitrary $\beta$, $d_\text{eff} = \text{Tr} \{
\exp[-\beta(H-E_0)] \}$ is the partition function with ground-state
energy $E_0$, reflecting the number of thermally occupied states, and
also scales exponentially fast with $L$ \cite{steinigeweg14}.
Therefore, while the error is exactly zero in the thermodynamic limit
$L \to \infty$, this error can be already very small at finite but
large $L$ and sampling is unnecessary, as is the case for all examples
considered in our work.

\subsection{Numerical implementation}

Most importantly, the approximation in Eq.\ (\ref{approximation1}) can
be calculated without knowing the eigenstates and eigenvalues of the
Hamiltonian. This calculation is based on the two auxiliary pure states
\begin{eqnarray}
&& | \Phi_N(\beta,t) \rangle = e^{-\imath H t -\beta H/2} \, |
\psi_N \rangle \, ,  \label{state1} \\
&& | \varphi_N(\beta,t) \rangle = e^{-\imath H t} \, j \, e^{-\beta H/2}
\, |\psi_N \rangle \, .
\label{state2}
\end{eqnarray}
Both states are time- and temperature-dependent and the only difference
between the two states is the additional current operator $j$ in the
r.h.s.\ of Eq.\ (\ref{state2}). Using these states, the approximation in
Eq.\ (\ref{approximation1}) reads
\begin{equation}
C(t) = \frac{\text{Re} \, \sum_N \langle \Phi_N(\beta,t)
| j | \varphi_N(\beta,t) \rangle}{L \, \sum_N \langle \Phi_N(\beta,0) |
\Phi_N(\beta,0) \rangle} \, . \label{approximation2}
\end{equation}
Apparently, the full time and temperature dependence in Eq.\
(\ref{approximation2}) results from the  evolution of the pure states only, 
i.e., there the current operator
$j$ is simply applied to the initial or time-evolved states.

For, e.g., $| \Phi_N(\beta,t)\rangle$, the $\beta$ dependence
is generated by an imaginary-time Schr\"odinger equation,
\begin{equation}
\imath \, \frac{\partial}{\partial (\imath \beta)} \, |
\Phi_N(\beta,0) \rangle = \frac{H}{2} \,  | \Phi_N(\beta,0)
\rangle \, , \label{imagS}
\end{equation}
and the $t$ dependence by the usual real-time Schr\"odinger
equation,
\begin{equation}
\imath \, \frac{\partial}{\partial t} \, | \Phi_N(\beta,t)
\rangle = H \,  | \Phi_N(\beta,t) \rangle \, . \label{realS}
\end{equation}
These differential equations can be solved by the use of straightforward
iterative methods such as, e.g.,\ Runge-Kutta \cite{elsayed2013, steinigeweg14,
steinigeweg14a}. We use a massively parallel implementation of a
Suzuki-Trotter product formula or Chebyshev polynomial algorithm
\cite{deraedt2007, jin2010}, allowing us to study quantum systems with as
many as $2 L=36$ lattice sites ($L=18$ in the fermionic language), where
the Hilbert-space dimension is $d = {\cal O}(10^{11})$. As compared to
exact diagonalization, this dimension is larger by orders of magnitude. Yet,
we do not exploit translation invariance of Hamiltonian and current. This
symmetry adds momentum as a good quantum number and an additional layer of
parallelization \cite{steinigeweg14a}.

In practice, we use the Chebyshev polynomial algorithm to compute $e^{-\beta
H/2}| \psi_N \, \rangle$. The results of this algorithm are exact to at least
ten digits. For the propagation in real time, we mostly use a unitary,
second-order product formula algorithm with a time step $\delta t \, t_\text{h}
= 0.02$, which is sufficiently small to guarantee that the total energy is
conserved up to at least six digits. Occasionally, we have used the Chebyshev
polynomial algorithm to compute the real-time evolution: no significant
differences between these and the product-formula results were found. Most of
the simulations were carried out on JUQUEEN, the IBM Blue Gene/Q located at
the J\"ulich Supercomputer Centre. A simulation of the largest system studied
in the present paper ($36$ spins) required $3$ TB of memory, the computation
was distributed over $131,072$ (MPI) processes, the total elapsed time to carry
out $400$ times steps was about $10$ hours ($1.3$ million core hours).

\section{Results for the Hubbard model}
\label{sec:hubbard}

This section contains our results for the charge transport in the 1D Hubbard
model, focusing at half filling. We consider infinite temperature
$\beta=1/T \to 0$ unless stated otherwise. First, we discuss the overall time
dependence of the current autocorrelation function for various values of $U/t_\text{h}$.
Second, we extract the Drude weight $D$ from the long-time behavior of $C(t)$.
Finally, we discuss the frequency dependence of the regular part and its
zero-frequency limit.

\subsection{Time dependence of autocorrelation functions}

\begin{figure}[tb]
\begin{center}
\includegraphics[width=0.9\columnwidth]{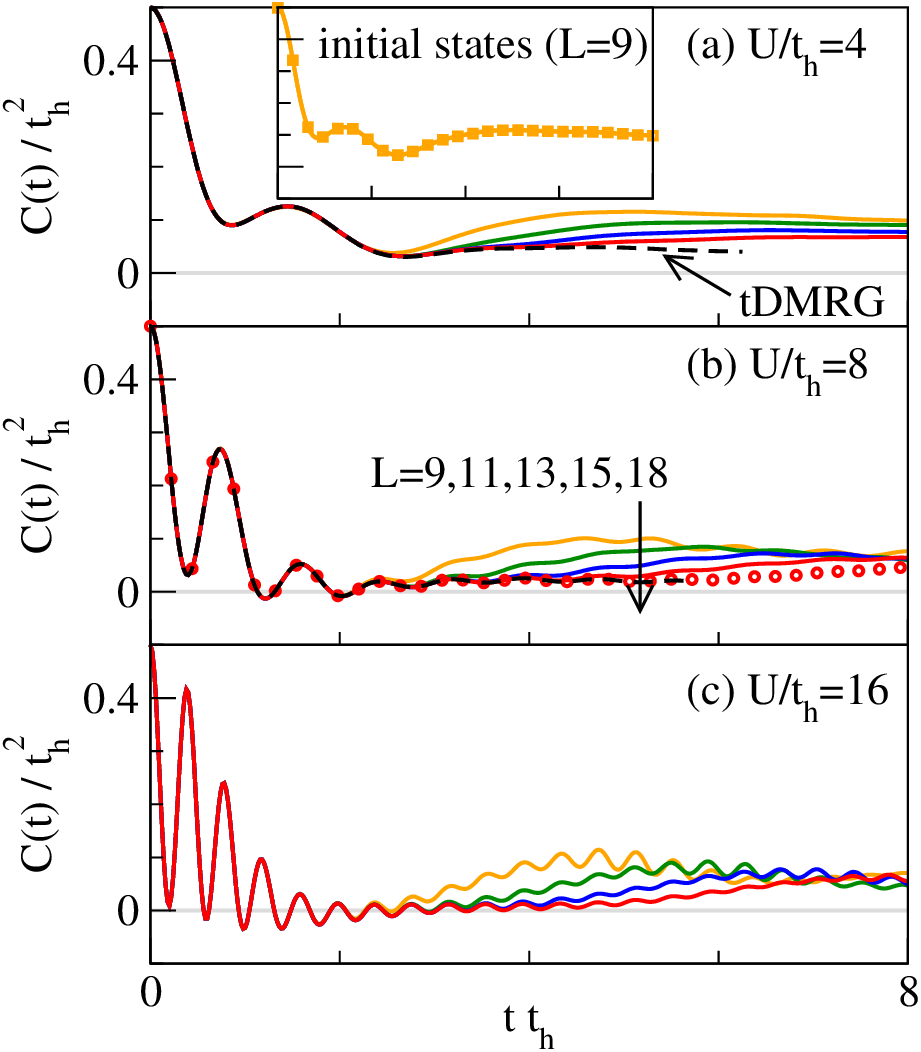}
\end{center}
\caption{(Color online) Real-time decay of the current autocorrelation function
$C(t)$ for (a) $U/t_\text{h} = 4$, (b) $U/t_\text{h} = 8$, (c) $U/t_\text{h} = 16$ for various $L = 9,
11, 13, 15, 18$ and at infinite  temperature $\beta \, t_\text{h} \to 0$ (solid curves
and circles). For the largest $L = 15$ and $18$, convergence to system-size
independent values is reached at times $t \, t_\text{h} \sim 5$. For comparison, tDMRG
data from \cite{karrasch14a} are included in (a) and (b) (dashed curves).
The inset in (a) shows the $L=9$ result for two different initial random
states (solid curve: first state, squares: second state), which demonstrates small
statistical errors for this $L$ already.}
\label{Fig1}
\end{figure}

Figures~\ref{Fig1}(a)-(c) show typical results for the real-time decay of the current
autocorrelation function $C(t)$ for  $U/t_\text{h} =4,8,16$, respectively, and several system
sizes $L\leq 18$. The figures show $C(t)$ for times up to $t \, t_\text{h} \lesssim 8$, where
the dominant decay of $C(t)$ from its initial value occurs. Typically, the data from
these different $L$ coincide for $t \, t_\text{h} \lesssim 2.5$. Beyond $t \, t_\text{h}=2.5$, $C(t)$
is a monotonically decreasing function of system size as indicated by the arrow in
Fig.~\ref{Fig1}(b). The figures further include real-time density matrix renormalization
group (tDMRG) data from \cite{karrasch14a} for comparison. Our DQT
results are in excellent agreement with the tDMRG data.

As $U/t_\text{h}$ increases, $C(t)$ approaches small values increasingly faster as a
function of time. On the other hand, the larger $U/t_\text{h}$, the more high-frequent
and pronounced are the oscillations in $C(t)$. These are inherited from the large
$U/t_\text{h}$ limit, in which the spectrum consists of bands of eigenstates separated
by gaps of order $U$. These bands correspond to excitations with multiple
doublons. Thus, the oscillatory dynamics in $C(t)$ at large $U/t_\text{h}$ is quite similar
to the behavior in the spin-1/2 XXZ chain in the strong Ising limit \cite{karrasch14}
and spin-1/2 XX ladders in the strong rung-coupling limit \cite{steinigeweg14a}.

\subsection{Drude weight}

\begin{figure}[tb]
\begin{center}
\includegraphics[width=0.9\columnwidth]{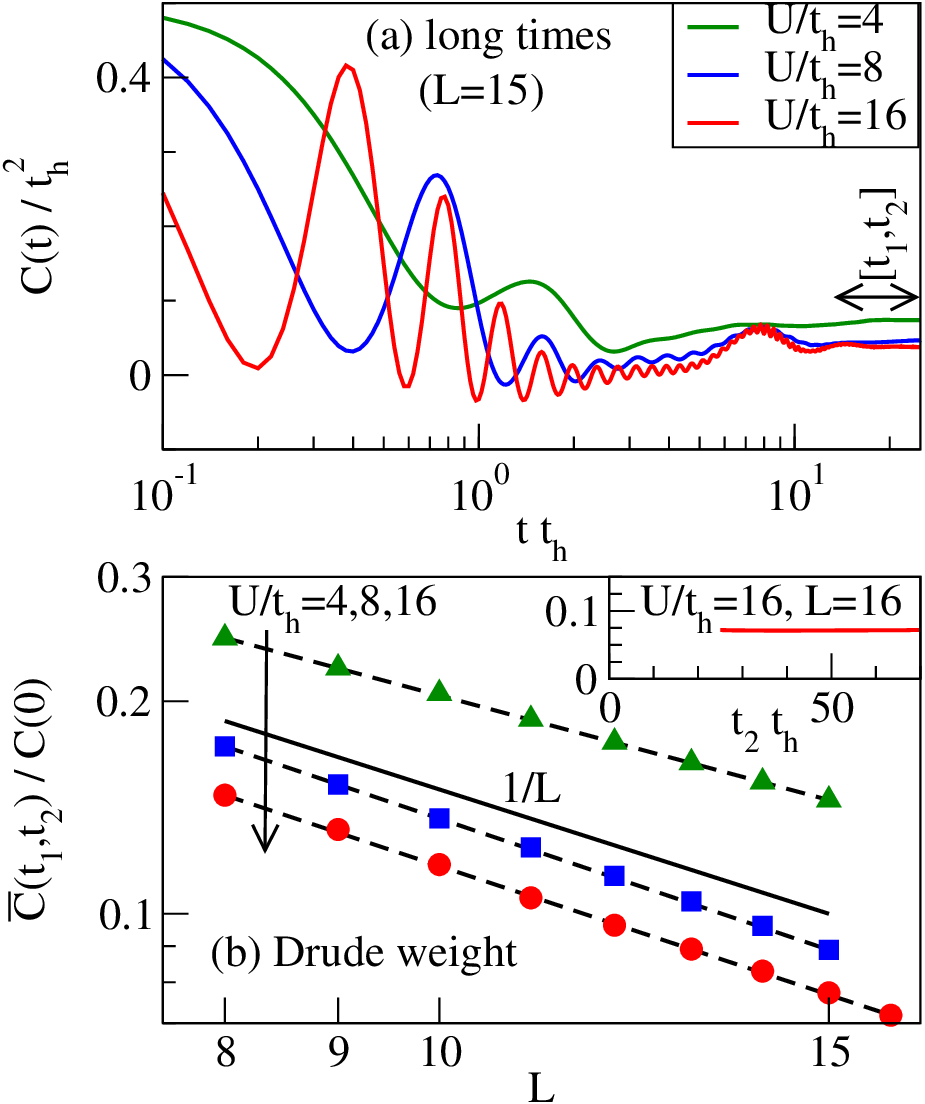}
\end{center}
\caption{(Color online) (a) Long-time limit of the current autocorrelation
function $C(t)$ for different $U/t_\text{h} = 4,8,16$, fixed $L = 15$, and high
temperatures $\beta \, t_\text{h} \to 0$. (b) Finite-size scaling of the Drude weight
$\bar{C}(t_1,t_2)$, as extracted from the time interval $[t_1 \, t_\text{h}, t_2 \,
t_\text{h}] = [12.5,25]$, in a log-log plot. As a guide to the eyes, power laws (dashed
lines) and a function $\propto 1/L$ (solid line) are indicated. The inset in
(b) shows, for $U/t_\text{h} = 16$ and $L=16$, that $\bar{C}(t_1,t_2)$ does not depend on
the specific choice of $t_2$.}
\label{Fig2}
\end{figure}

\begin{figure}[tb]
\begin{center}
\includegraphics[width=0.8\columnwidth]{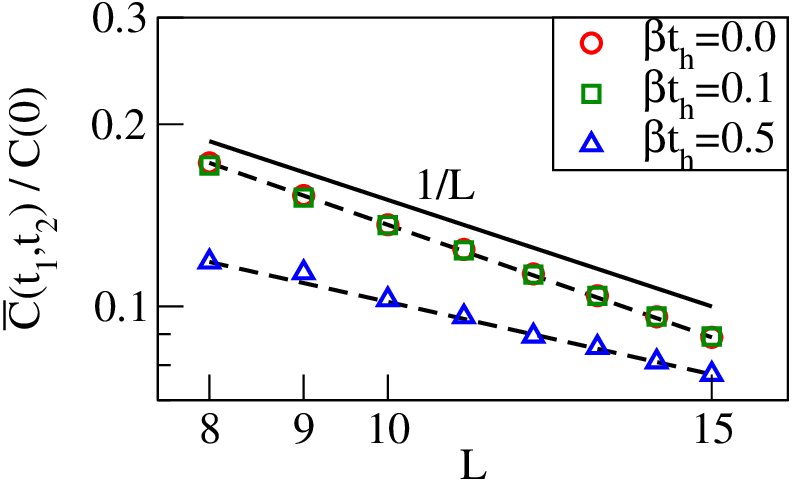}
\end{center}
\caption{(Color online) The same information as shown in Fig.\ \ref{Fig2}
(b) but for fixed $U / t_\text{h} = 8$ and different $\beta \, t_\text{h} = 0, 0.1, 0.5$.
Data for $\beta \, t_\text{h} = 0.0$ and $0.1$ almost coincide.}
\label{Fig3}
\end{figure}

In order to extract the non-decaying portion of $C(t)$, which  equals the
Drude weight, much longer times than $t \, t_\text{h} \sim 12$ need to be considered
\cite{steinigeweg14}. Therefore, we display $C(t)$ for $t \, t_\text{h} \leq 25$ in
Fig.~\ref{Fig2}(a) for the example of $U/t_\text{h}=4,8,16$ and for $L=15$. At times
$t \, t_\text{h} \geq 10$, the oscillations in $C(t)$ have decayed to a sufficiently
small amplitude and hence we estimate the Drude weight by averaging $C(t)$
in the time window $t \in [t_1 \, t_\text{h} =12.5,t_2 \, t_\text{h} =25]$, yielding
$\bar C(t_1,t_2)$. Note that $\bar C(t_1,t_2)$ does not depend on
this specific choice of $t_2$, see the inset of Fig.\ \ref{Fig2}(b).

The resulting, $L$-dependent $\bar C(t_1,t_2)$ are shown in Fig.~\ref{Fig2}(b)
in a log-log plot. The system-size dependence of $\bar C(t_1,t_2)$ is consistent
with a $1/L$ decay of the Drude weight to zero as system size increases. This
scaling of $D$ with system size is typical for integrable systems: it has been
observed for the spin Drude weight of the spin-1/2 XXZ chain as well \cite{hm03,
herbrych11,karrasch13,steinigeweg14}. Moreover, the Drude weight approximately
measures the fluctuations of diagonal matrix elements of the  associated current
operator \cite{steinigeweg13}. Such system-size dependent fluctuations are
commonly investigated to access the validity of the eigenstate thermalization
hypothesis \cite{deutsch91, srednicki94, rigol08}. For integrable systems, most
numerical studies indicate a slow, power-law decay of these fluctuations
\cite{steinigeweg13,beugeling14,alba15}. Most notably, our data are consistent
with a vanishing Drude weight $D = 0$ at infinite temperature, in
agreement with \cite{carmelo12}.

In principle, if the infinite-temperature Drude weight vanishes, this does not
necessarily imply that $D(T)=0$ at any finite $T$. To see this, one can write
the Drude weight in a high-temperature expansion
\begin{equation}
D(T) = \frac{D_1}{T} + \frac{D_2}{T^2} + \dots
\end{equation} 
where $D_1$ is the infinite-temperature Drude weight studied in Fig.~\ref{Fig2}(b).
To substantiate that in the Hubbard model at half filling $D(T)=0$ at any finite $T$,
we have also computed $D(T)$ at $T/t_\text{h}= 2,10$, where $D$ also seems to vanish as $L$
increases. This is illustrated in Fig.\ \ref{Fig3}.

\subsection{Optical conductivity}

\begin{figure}[tb]
\begin{center}
\includegraphics[width=0.9\columnwidth]{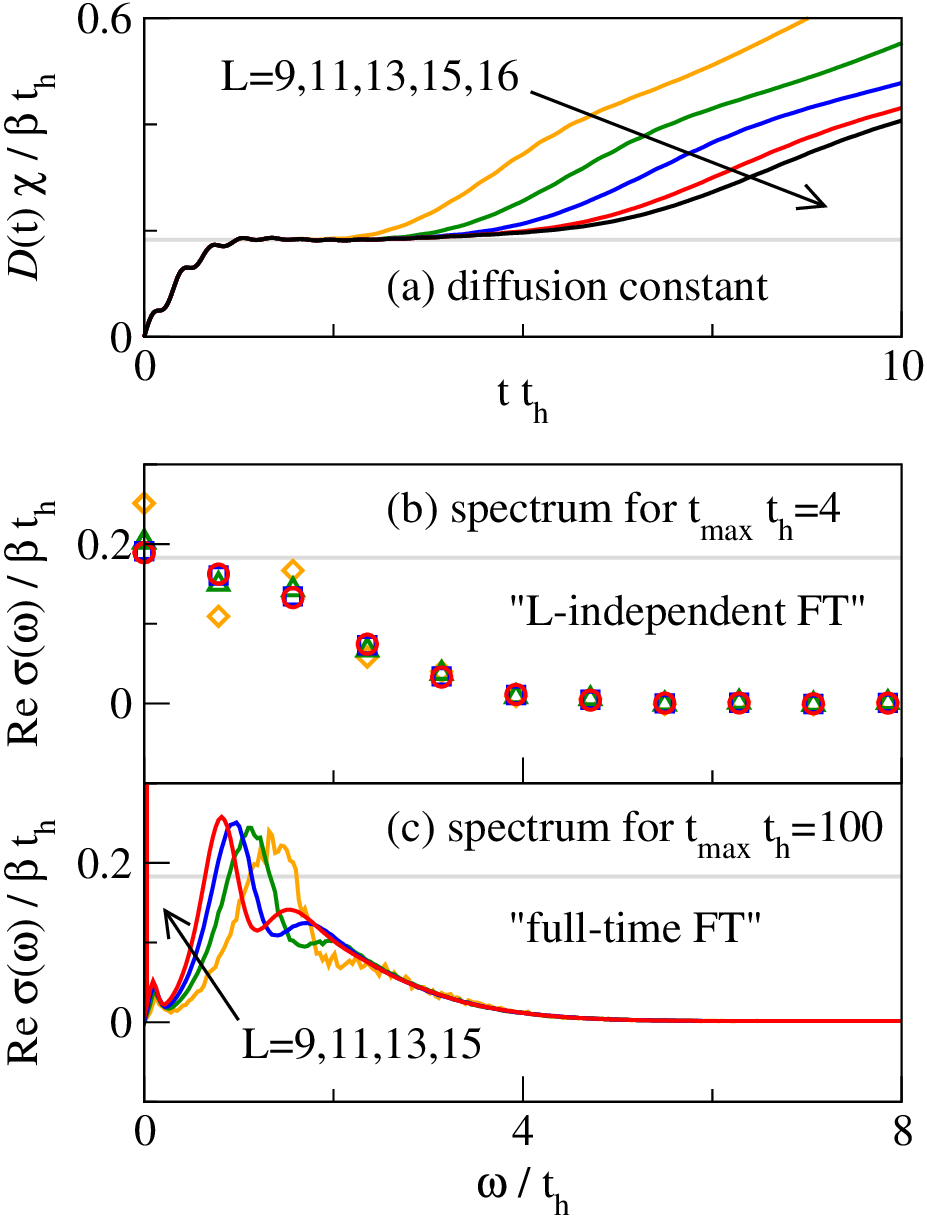}
\end{center}
\caption{(Color online) (a) Time-dependent diffusion constant ${\cal D}(t)$ for
$U/t_\text{h} = 16$, various $L = 9,11,13,15,16$, and high temperatures $\beta \, t_\text{h} \to
0$. A plateau is clearly visible at intermediate times before the finite-size Drude
weight yields a linear increase in the long-time limit. The plateau height is independent
of $L$ and the plateau width increases with $L$. (This behavior is almost identical
to the XXZ chain at $\Delta > 1$.) (b), (c) Frequency dependence of the optical
conductivity $\text{Re} \, \sigma(\omega)$, as resulting from $t_\text{max} \, t_\text{h}=4,
100$. (c) does not respect the ``better'' limit of $L \to \infty$ first and $t_\text{max}
\, t_\text{h} \to \infty$ afterward. Apparently, (c) shows strong finite-size effects at both,
$\omega = 0$ and $\omega \neq 0$. However, in the thermodynamic limit $L \to \infty$,
(c) seems to approach (b).}
\label{Fig4}
\end{figure}

\begin{figure}[tb]
\begin{center}
\includegraphics[width=0.8\columnwidth]{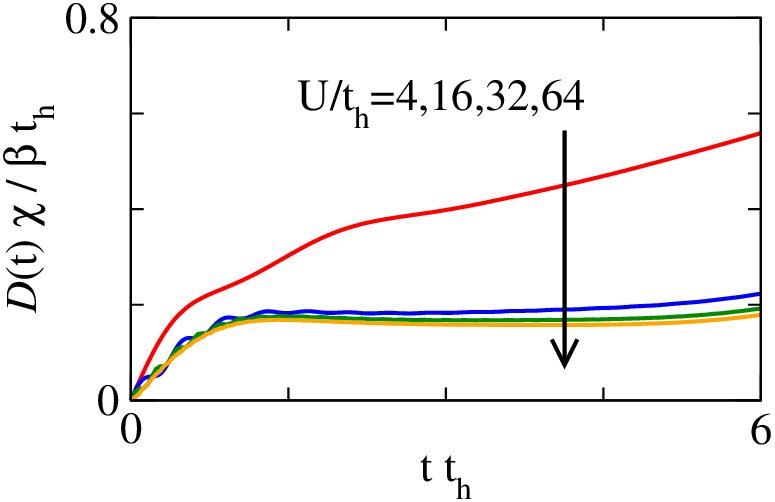}
\end{center}
\caption{(Color online) Time-dependent diffusion constant ${\cal D}(t)$ for
various $U/t_\text{h} = 4,16,32,64$, fixed $L = 15$, and high temperatures $\beta \, t_\text{h}
\to 0$. Clearly, the plateau value of ${\cal D}(t)$ becomes independent of $U$
in the limit of large $U$.}
\label{Fig5}
\end{figure}

\begin{figure}[tb]
\begin{center}
\includegraphics[width=0.9\columnwidth]{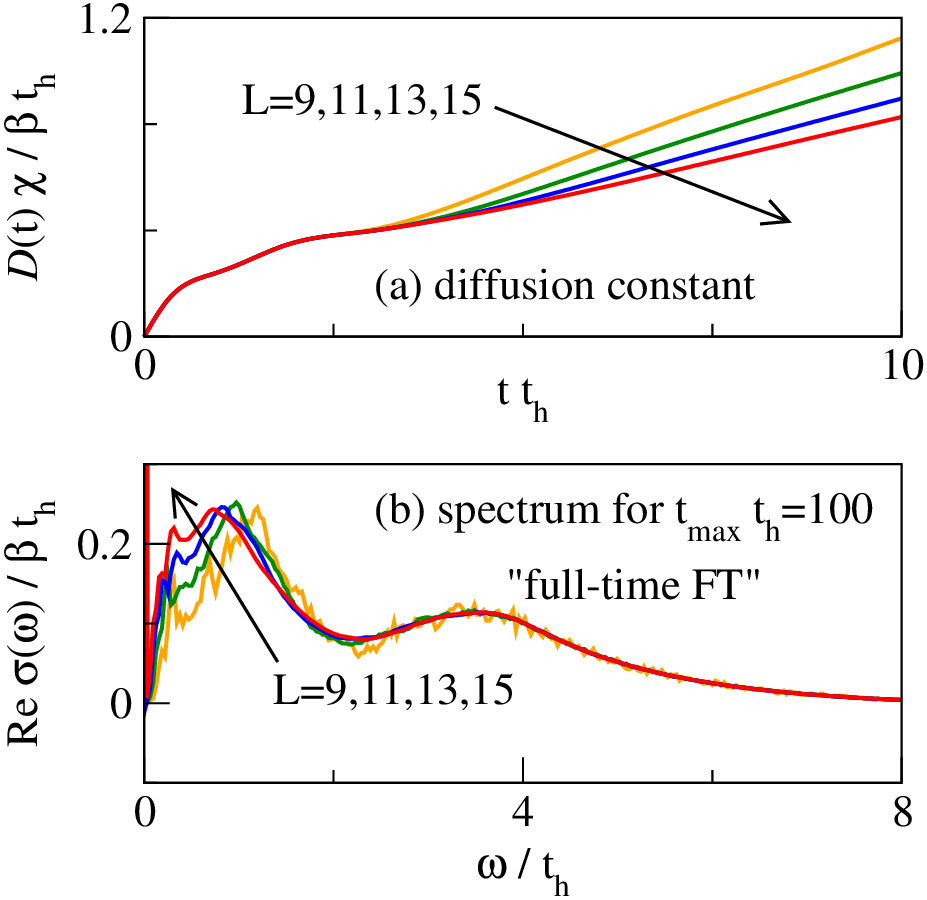}
\end{center}
\caption{(Color online) The same information as shown in Fig.~\ref{Fig4}, yet here
for $U/t_\text{h} = 4$. Extracting $\text{Re} \, \sigma(\omega)$ via the $L$-independent FT
strategy is not applicable here, since $\mathcal{D}(t)$ does not exhibit a clear
plateau [see (a)], due to the long-time tail in $C(t)$. This plateau will also not
occur for $L = 18$. Thus, in (b), we present $\text{Re} \, \sigma(\omega)$ obtained
from a full-time FT, which thus results in strong finite-size effects at small
frequencies.}
\label{Fig6}
\end{figure}

Since the Drude weight appears to vanish as $L\to \infty$, all weight in
$\text{Re} \, \sigma(\omega)$ will ultimately be in the regular part
$\sigma_{\rm reg}(\omega)$. This optical conductivity has recently been
studied using tDMRG \cite{karrasch14a}, where a finite dc conductivity was
observed that diverges as $\sigma_{\rm dc} \sim 1/T$ as temperature decreases.

We here first demonstrate that it is indeed possible to extract the dc
conductivity from our time-dependent data for $C(t)$. At infinite temperature,
the dc conductivity $\sigma_\text{dc}/\chi$ is simply equal to the
integral $\mathcal{D}(t)$ over $C(t)$ as defined in Eq.~\eqref{eq:diff}, i.e.,
connected to the diffusion constant by an Einstein relation.

At large $U/t_\text{h}$, $\mathcal{D}$ increases quickly and then settles into a plateau,
as is evident from the example presented in Fig.~\ref{Fig4}(a). At large times,
$\mathcal{D}$ further increases, which is due to both the non-zero Drude weight on
finite systems and other finite-size effects. Plotting data for $\mathcal{D}$ for
several system sizes clearly suggests that finite-size data gradually approach
the plateau value at longer times as well, see Fig.~\ref{Fig4}(a).

The presence of such a plateau, following the reasoning of \cite{steinigeweg09},
suggests a finite dc conductivity and diffusion constant. As shown in Fig.\
\ref{Fig5}, the diffusion constant exhibits a peculiar behavior at
$T=\infty$: As $U/t_\text{h}$ increases, it saturates at a {\it $U$-independent} value.
This saturation results from the structure of the energy spectrum in the
large-$U/t_\text{h}$ limit: It consists of bands separated by $U$ that have a band width
given by $t_\text{h}$. Since we are taking the limit $U/t_\text{h}\to \infty$ {\it after} taking
the limit $T\to \infty$, the dominant contribution to scattering comes from interband
processes. This behavior appears to be generic for systems with an emergent
ladder-like spectrum and has also been observed in the Ising regime of spin-$1/2$
XXZ chains \cite{karrasch14} and in spin-1/2 XX ladders \cite{steinigeweg14a}.
The independence of the diffusion constant on $U$ observed in Fig.\ \ref{Fig5}
also unveils that the lower bound of \cite{prosen14}, as given in Eq.\
(\ref{lowerbound}), is not exhaustive in the large-$U/t_\text{h}$ regime.

For the purpose of computing $\text{Re} \, \sigma(\omega)$, the existence of
the plateau implies that the asymptotic behavior has been reached. Moreover, the value of
the plateau in $\mathcal{D}(t)$ is independent of system size for the parameters
of Fig.~\ref{Fig4}. Thus, we will compare two ways of computing $\text{Re} \,
\sigma(\omega)$: (i) the first version uses the full time dependence of $C(t)$, up to
and including times where we clearly observe finite-size effects (later dubbed full-time FT);
(ii) In the second, we restrict the time window for the Fourier transformation  to
times at which we have system-size independent data for $C(t)$ (later referred to
as $L$-independent FT).

The results of both approaches are presented in Figs.~\ref{Fig4}(b) and (c),
respectively. The full-time FT resolves the strong finite-size dependent structures
that were known to exist from Ref.~\cite{prelovsek04}. The positions of these sharp
peaks shift to smaller frequencies as system size increases. An extrapolation of
$\,\sigma_{\rm reg}(\omega)$ to zero frequency is thus difficult to control.

The behavior of $\, \sigma_{\rm reg}(\omega)$ computed using the $L$-independent
FT strategy, by contrast, is a very smooth function that strongly resembles the
optical conductivity of a typical diffusive system. This is clearly related to the
fast initial decay of $C(t)$ [see the data shown in Fig.~\ref{Fig1}(b)], and the 
corresponding establishment of the plateau in the integrated quantity
$\mathcal{D}(t)$, which consequently allows us to estimate the dc limit under the
assumption that no additional time dependence emerges in $C(t)$ at very long times
and large systems. We thus propose that whenever such a plateau is present in
$\mathcal{D}(t)$, the cleanest way of computing $ \sigma_{\rm reg}(\omega)$
is the  $L$-independent FT, in line with the reasoning of Refs.~\cite{steinigeweg09,
karrasch14,karrasch15}.

Figure~\ref{Fig6} shows data for $U/t_\text{h}=4$ as an example for a case, in which no
clear plateau in $\mathcal{D}(t)$ can be resolved with the accessible system
sizes. Here, we thus compute $\, \sigma_{\rm reg}(\omega)$ from the full available
time series of $C(t)$, which is shown in Fig.~\ref{Fig6}(b). The optical conductivity
has a broad maximum at $\omega/t_\text{h} \sim U/t_\text{h}$ and an additional low-frequency peak
at $\omega /t_\text{h} \sim 1$ whose position shifts to small frequencies as $L$
increases. The data would suggest a small or vanishing dc conductivity, which we
believe does not reflect the behavior of an infinitely large system [compare
Fig.~\ref{Fig4}(b)], since the low-frequency finite-size effects likely screen
the correct low-frequency dependence.

\section{Results for the mass-imbalanced case}
\label{sec:mass}

In this section, we present our results for the mass-imbalanced cases $\eta = 
t_\downarrow / t_\uparrow < 1$, where the model is non-integrable. We start
with the case $\eta = 0$, the Falicov-Kimball limit, and discuss the emergence
of  Anderson localization in this limit. Then we turn to the case of
$\eta \sim 1/2$ and study both, Drude weight and optical conductivity.
Finally, we summarize the scaling of the diffusion constant as a function of
$\eta$ in the $\eta$ region accessible to our numerical method.

\subsection{Falicov-Kimball limit}

\begin{figure}[tb]
\begin{center}
\includegraphics[width=0.9\columnwidth]{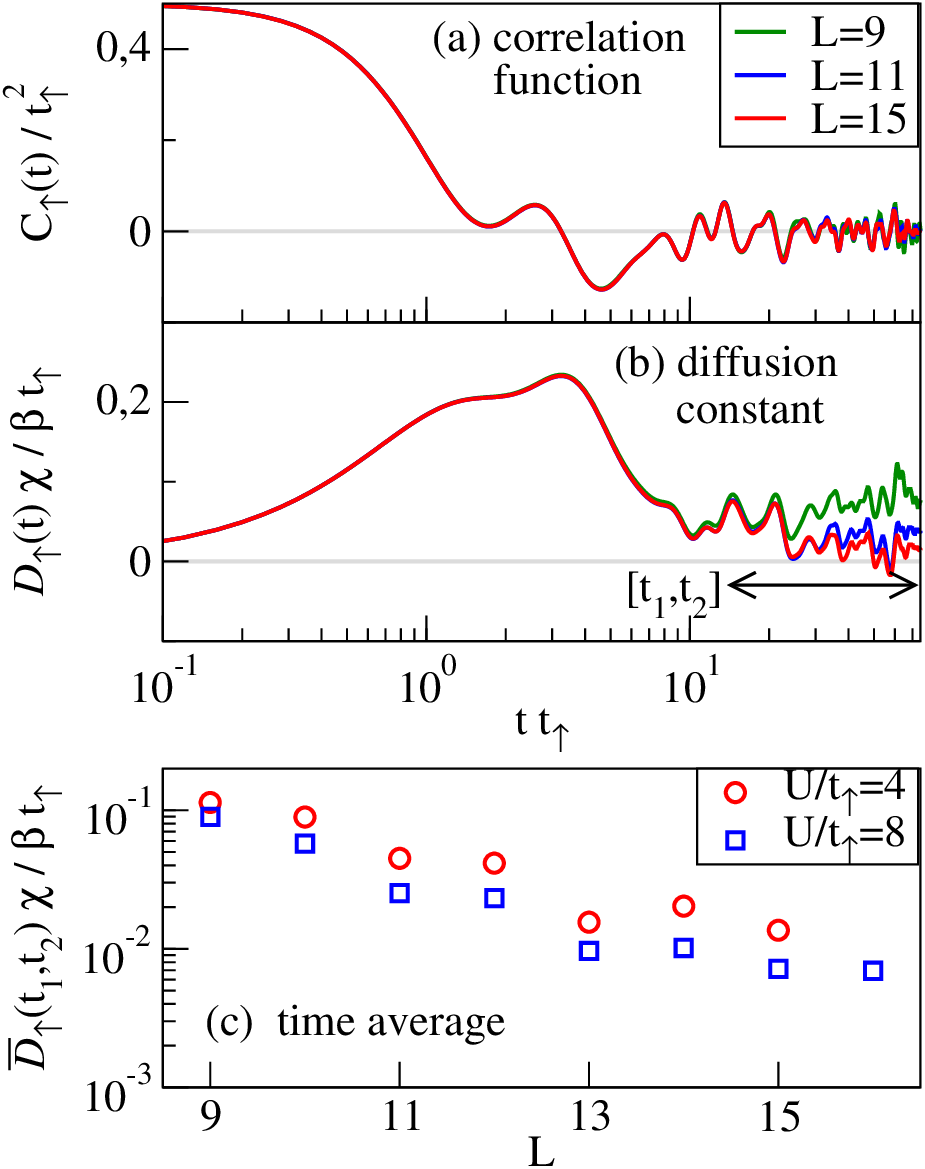}
\end{center}
\caption{(Color online) (a) Real-time decay of the current autocorrelation
function $C_\uparrow(t)$ of the light $\uparrow$-component for $U/t_\uparrow = 4$,
strong imbalance $\eta = t_\downarrow/t_\uparrow = 0$, $L = 9, 11, 15$, and high
temperatures $\beta \, t_\uparrow \to 0$. Since $C_\uparrow(t)$ is highly
oscillating after the initial decay, also the time-dependent diffusion
constant ${\cal D}_\uparrow(t)$ does so in (b). Consequently, the usual extraction
of a diffusion constant would depend on the specific point in time considered.
However, the time average still yields a reasonable diffusion constant. (c)
Finite-size scaling of the time average for $U/t_\uparrow=4,8$, as resulting from
the time interval $[t_1 \, t_\uparrow, t_2 \, t_\uparrow] = [12.5,75$], in a
semi-log plot. Apparently, the scaling is non-trivial, but the decrease is
consistent with insulating behavior in the thermodynamic limit $L \to \infty$.}
\label{Fig7}
\end{figure}

\begin{figure}[tb]
\begin{center}
\includegraphics[width=0.9\columnwidth]{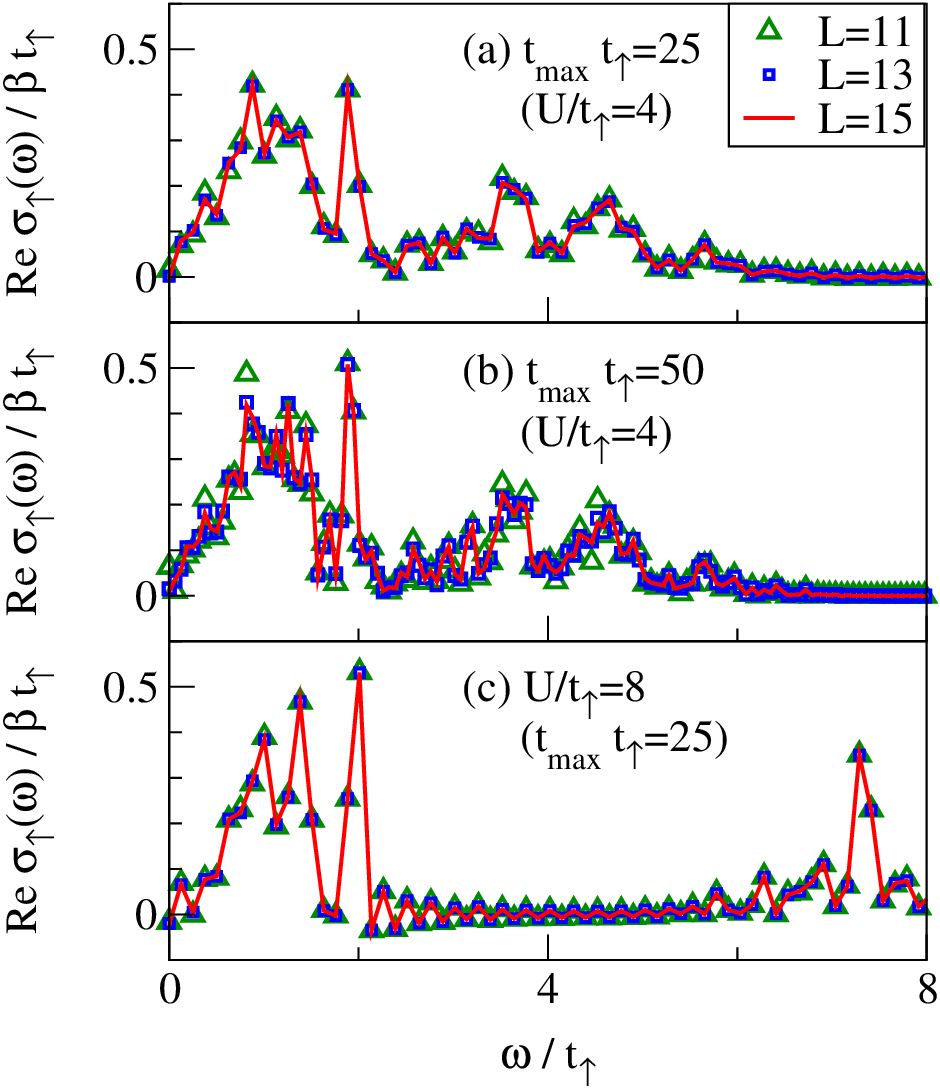}
\end{center} 
\caption{(Color online) Frequency dependence of the optical conductivity
$\text{Re} \, \sigma_\uparrow(\omega)$ for (a) $t_{\rm max} \, t_\uparrow = 25$,
(b) $t_{\rm max} \, t_\uparrow = 50$ for $U/t_\uparrow = 4$, strong imbalance
$\eta = t_\downarrow /t_\uparrow = 0$, different $L = 11,13,15$, and high
temperatures $\beta \, t_\uparrow \to 0$. (c) shows (a) for $U/t_\uparrow = 8$.
The overall structure is independent of $t_{\rm max}$ and $L$. For the dependence
of the dc limit $\text{Re} \, \sigma_\uparrow(\omega \to 0)$ on $L$ and
$t_\text{max}$, see $\bar{D}_\uparrow(t_1,t_2)$ discussed before.} 
\label{Fig8} 
\end{figure}

In the Falicov-Kimball limit $\eta = 0$ the model simplifies to
\begin{equation}
h_l= -t_\uparrow \left(c_{l,\uparrow}^\dagger c_{l+1,\uparrow}^{\phantom{\dagger}}
+ \tn{h.c.} \right) + {U}(n_{l,\uparrow}-\frac{1}{2})(n_{l,\downarrow}-\frac{1}{2}) 
\, .
\end{equation}
For this simplified model all $n_{l,\downarrow}$ commute with all local
Hamiltonians $h_l$ and with each other,
\begin{equation}
[h_l, n_{k,\downarrow} ] = [n_{l,\downarrow},n_{k,\downarrow}] = 0 \, ,
\end{equation}
$l,k = 1, \ldots, L$. Each $(n_{l,\downarrow} -1/2)$ is thus conserved
and yields a good quantum number $\epsilon_l = \pm 1/2$, with $2^L$ different
sequences
\begin{equation}
\epsilon(m) = (\epsilon_1(m), \ldots, \epsilon_L(m)) \, ,
\end{equation}
$m = 1, \ldots, 2^L$. As a consequence, the full Hamiltonian $H = \sum_l h_l$ can 
be rewritten as a sum of $2^L$ uncoupled Hamiltonians $H(m) = \sum_l h_l(m)$,
where
\begin{equation}
h_l(m) = -t_\uparrow \left(c_{l,\uparrow}^\dagger c_{l+1,\uparrow}^{\phantom{\dagger}}
+ \tn{h.c.} \right) + U \, \epsilon_l(m) \, (n_{l,\uparrow}-\frac{1}{2})
\end{equation}
and the $U$ part becomes a site-dependent potential given by the
sequence $\epsilon(m)$. For many $m$, $\epsilon(m)$ can be
understood as a sequence of random numbers drawn from a binary distribution
$[-1/2,1/2]$. Therefore, remarkably, many uncoupled Hamiltonians $H(m)$ can
be interpreted also as the single-particle, Anderson problem for on-site
disorder of strength $U$. Note that translation invariance is typically
broken for a given $m$ but restored by sampling over $m$. Note further
that {\it all} $m$ contribute at finite temperatures.

Due to the analogy to the single-particle, Anderson problem and the strict
one-dimensionality of the lattice, one expects perfectly insulating behavior
in the thermodynamic limit $L \to \infty$ at all temperatures. Early on, this
expectation has been verified in numerical calculations of the optical conductivity
\cite{devries93,devries94} for $\beta \, t_\uparrow > 0$ and values of $U$
where the localization length does not exceed lattice sizes accessible.
Yet, the high-temperature limit $\beta \, t_\uparrow \to 0$ has not been
studied.

In Fig.~\ref{Fig7}(a) we show our results for the time-dependent current
autocorrelation function $C_\uparrow(t)$ for $\beta \, t_\uparrow \to 0$, $U /
t_\uparrow = 4$, and different $L = 9, 11, 15$. Clearly, $C_\uparrow(t)$ decays
rapidly on a rather short time scale $t \, t_\uparrow \sim 1$. After this initial
decay $C_\uparrow(t)$ approaches zero from the negative side but still shows small
oscillations. Note that these oscillations are no finite-size effects since curves
for $L = 11$ and $15$ are practically identical to each other for the long times
$t \, t_\uparrow \sim 75$ depicted in the figure. This curve for $C_\uparrow(t)$
yields the time-dependent diffusion constant ${\cal D}_\uparrow(t)$ shown in
Fig.~\ref{Fig7}(b). After the initial increase of ${\cal D}_\uparrow(t)$ we
find a strong decrease related to the region where $C_\uparrow(t)$ is negative.
Necessarily, ${\cal D}_\uparrow(t)$ also shows small oscillations not related
to finite-size effects, as evident from comparing $L=11$ and $15$ again.

The long-time oscillations of ${\cal D}_\uparrow(t)$ indicate that the dynamical
process cannot be described by a diffusion constant in the strict sense. However, to
extract an effective diffusion constant, we average ${\cal D}_\uparrow(t)$ over the
long-time interval $[t_1 \, t_\uparrow, t_2 \, t_\uparrow] = [12.5, 75]$. In
Fig.~\ref{Fig7}(c) we depict the resulting $\bar{\cal D}_\uparrow(t_1,t_2)$ as a
function of $L$ for $U / t_\uparrow = 4, 8$ in a semi-log plot. Apparently, this
time-averaged quantity decreases as system size increases and may eventually
become zero in the thermodynamic limit $L \to \infty$. Note that the scaling
for small $L$ is partially related to tiny finite-size Drude weights $D_\uparrow$,
entering ${\cal D}_\uparrow(t)$ via the relation ${\cal D}_\uparrow(t) \propto
D_\uparrow \, t$ in the long-time limit.

Next we turn to the optical conductivity. Since $C_\uparrow(t)$ and
${\cal D}_\uparrow(t)$ do not become constant in the long-time limit, the
finite-time Fourier transform necessarily depends on the specific time interval
chosen. Thus, we show in Figs.~\ref{Fig8}(a) and (b) the Fourier transform of
$U/t_\uparrow = 4$ data for $t_{\rm max} \ t_\uparrow = 25$ and $50$, where times
$t\leq t_{\rm max}$ where considered in the Fourier transformation. While
Figs.~\ref{Fig8}(a) and (b) differ with respect to  details, the overall structure
does not depend on the specific choice of $t_\text{max}$. In particular,
the limit $\omega \to 0$ is
consistent with a vanishing dc conductivity. Note that this limit coincides with
${\cal D}_\uparrow(t)$ evaluated at $t \, t_\uparrow = 25$ and $50$, respectively.
Similarly, our results indicate a vanishing dc conductivity for $U/t_\uparrow =
8$, as shown in Fig.~\ref{Fig8}(c). The small negative spectral weight is
an artifact of the finite-time Fourier transform used and depends on the specific
choice of $t_\text{max}$.

To summarize, our $\beta \, t_\uparrow \to 0$ results are consistent with the
interpretation of the model in terms of the single-particle, Anderson problem
in one spatial dimension.

\subsection{Intermediate imbalance}

\begin{figure}[tb]
\begin{center}
\includegraphics[width=0.9\columnwidth]{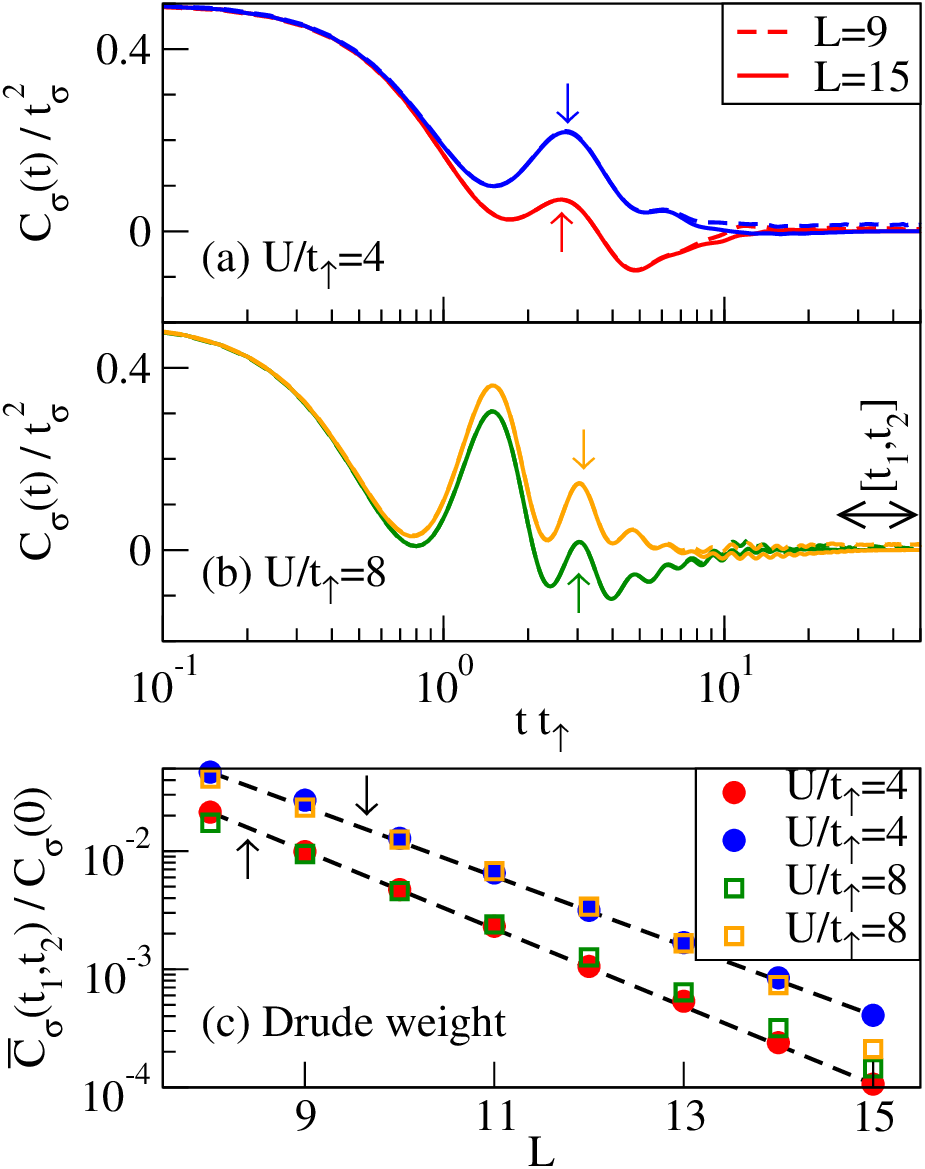}
\end{center}
\caption{(Color online) Real-time decay of the current autocorrelation function
$C_\sigma(t)$ for (a) $U/t_\uparrow = 4$, (b) $U/t_\uparrow = 8$ for $\eta =
t_\downarrow / t_\uparrow = 0.4$, both components $\sigma=\uparrow,\downarrow$,
two $L = 9, 15$, and high temperatures $\beta \, t_\uparrow \to 0$. (c) Finite-size
scaling of the Drude weight $\bar{C}_\sigma(t_1,t_2)$, as extracted from the time
interval $[t_1 \, t_\uparrow, t_2 \, t_\uparrow] = [25,50]$, in a semi-log plot.
As a guide to the eyes, exponentials (dashed lines) are indicated.}
\label{Fig9}
\end{figure}

\begin{figure}[tb]
\begin{center}
\includegraphics[width=0.9\columnwidth]{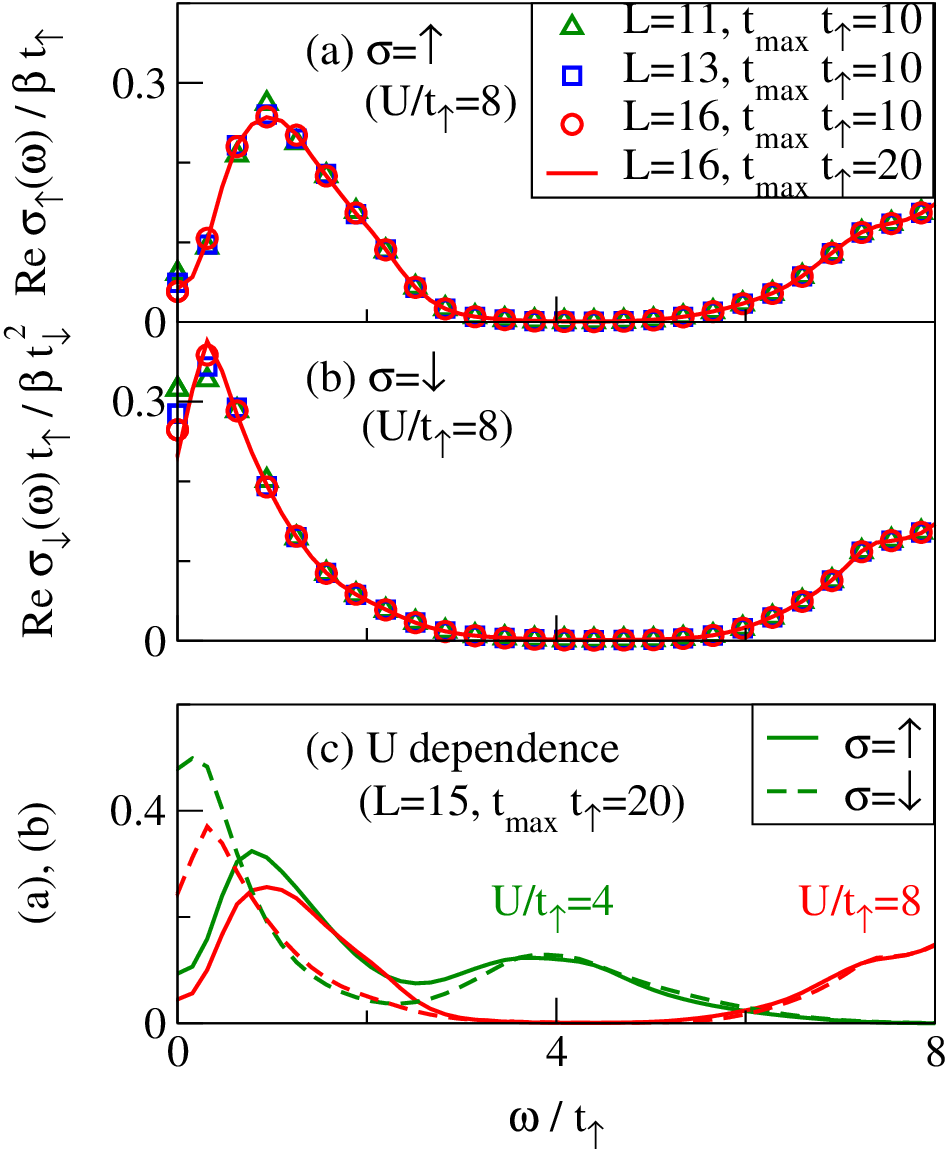}
\end{center}
\caption{(Color online) Frequency dependence of the {optical conductivity}
$\text{Re} \, \sigma_\sigma(\omega)$ for the (a) light component $\sigma=\uparrow$,
(b) heavy component $\sigma=\downarrow$ for $U/t_\uparrow = 8$, $\eta=  t_\downarrow
/t_\uparrow = 0.4$, and high temperatures $\beta \, t_\uparrow \to 0$, as resulting
from different $L = 11,13,16$ and $t_\text{max} \, t_\uparrow = 10,20$. The
independence of $L$ and {$t_\text{max}$} is evident. (c) $U$ dependence of
{$\text{Re} \, \sigma_\sigma(\omega)$} for a large $L = 15$ and long
${t_\text{max}} \, t_\uparrow = 20$. A peak at $\omega / t_\uparrow
=U/t_\uparrow$ is clearly visible.}
\label{Fig10}
\end{figure}

Next we discuss the region $0 < \eta < 1$, where the model still is
non-integrable but the interpretation of the model in terms of the
single-particle, Anderson problem is not possible any more. In fact,
in this $\eta$ region, we deal with a many-particle problem.

We start with intermediate imbalance $\eta = 0.4$. In Fig.~\ref{Fig9}(a)
we depict our results for the time-dependent current autocorrelation
function $C_\sigma(t)$ for the light  ($\sigma = \uparrow$) and the
heavy ($\sigma = \downarrow)$ component for $U / t_\uparrow = 4$ and
$L = 9,15$, still in the high-temperature limit $\beta \, t_\uparrow \to 0$.
In Fig.~\ref{Fig9}(b) we additionally show results for $U / t_\uparrow
= 8$. For both components, $C_\sigma(t)$ decays fast on a time scale $t \,
t_\uparrow \sim 1$ but revivals appear afterward. While these revivals
are equally pronounced for $\sigma = \uparrow$ and $\downarrow$, only
$C_\uparrow(t)$ becomes negative in the time interval $t \, t_\uparrow
\sim 2.5$. However, any revivals eventually disappear and $C_\sigma(t)$
decays fully to approximately zero for $\sigma = \uparrow$ and $\downarrow$.
When comparing curves for $L=9$ and $15$, it is also evident that finite-size
effects are small on the physically relevant time scale. Thus, we are able
to obtain information on $C_\sigma(t)$ in the thermodynamic limit $L \to
\infty$ without invoking intricate extrapolations.

It is also evident from Figs.~\ref{Fig9}(a) and (b) that Drude weights
$D_\sigma$ are small, i.e., there is no long-time saturation of $C_\sigma(t)$
at a significant positive value. However, it is instructive to discuss the
actual value of the Drude weights in more detail. In Fig.~\ref{Fig9}(c) we
show the finite-size scaling of $\bar{C}_\sigma(t_1,t_2)$, as extracted
from the time interval $[t_1 \, t_\uparrow, t_2 \, t_\uparrow] = [25,50]$,
for $\sigma = \uparrow, \downarrow$ and $U / t_\uparrow = 4, 8$ in a semi-log
plot. Interestingly, $\bar{C}_\sigma$ is larger for $\sigma =
\downarrow$ and does not depend on $U$. In all cases, the finite-size scaling
of $\bar{C}_\sigma$ is remarkably well described by a simple exponential
decrease over three orders of magnitude, with a relative value
$\bar{C}_\sigma / C_\sigma(0) < 10^{-3}$ at $L=15$. This exponential
decrease is expected for strongly non-integrable models \cite{hm04,steinigeweg14a}
and, moreover, is in accord with the eigenstate thermalization hypothesis
\cite{beugeling14,steinigeweg13}.

Since finite-size effects are small and $C_\sigma(t)$ decays to approximately
zero, we can accurately determine the optical conductivity by Fourier
transforming data for finite $L$ and $t$. In Figs.~\ref{Fig10}(a) and (b) we show the
finite-time optical conductivity $\text{Re} \, \sigma_\sigma(\omega)$ at
$U/t_\uparrow = 8$ for the light and heavy component, respectively. As expected,
{$\text{Re} \, \sigma_\sigma(\omega)$} does neither depend on
{$t_\text{max}$} nor $L$ and is a smooth function of frequency $\omega$. Similarly
to the integrable case $\eta = 0$, we find a broad maximum at $\omega / t_\uparrow
\sim U / t_\uparrow$ for both $\sigma$. In contrast, the position of the additional
peak at low $\omega$ depends on $\sigma$ but is roughly independent of $U$, as shown
in Fig.~\ref{Fig10}(c). Most importantly, the dc conductivity is finite and its
actual value is, relative to the amplitude of the low-$\omega$ peak, larger
for the {heavy component $\sigma = \downarrow$}. As a function of $U$, this dc
conductivity decreases but is still finite for all $U$ depicted, see
Fig.~\ref{Fig10}(c). Therefore, at $\eta = 0.4$, we can exclude the existence of an
insulator in the high-temperature limit $\beta \, t_\uparrow \to 0$.

\subsection{Scaling of diffusion constant and dc conductivity}

\begin{figure}[tb]
\begin{center}
\includegraphics[width=0.9\columnwidth]{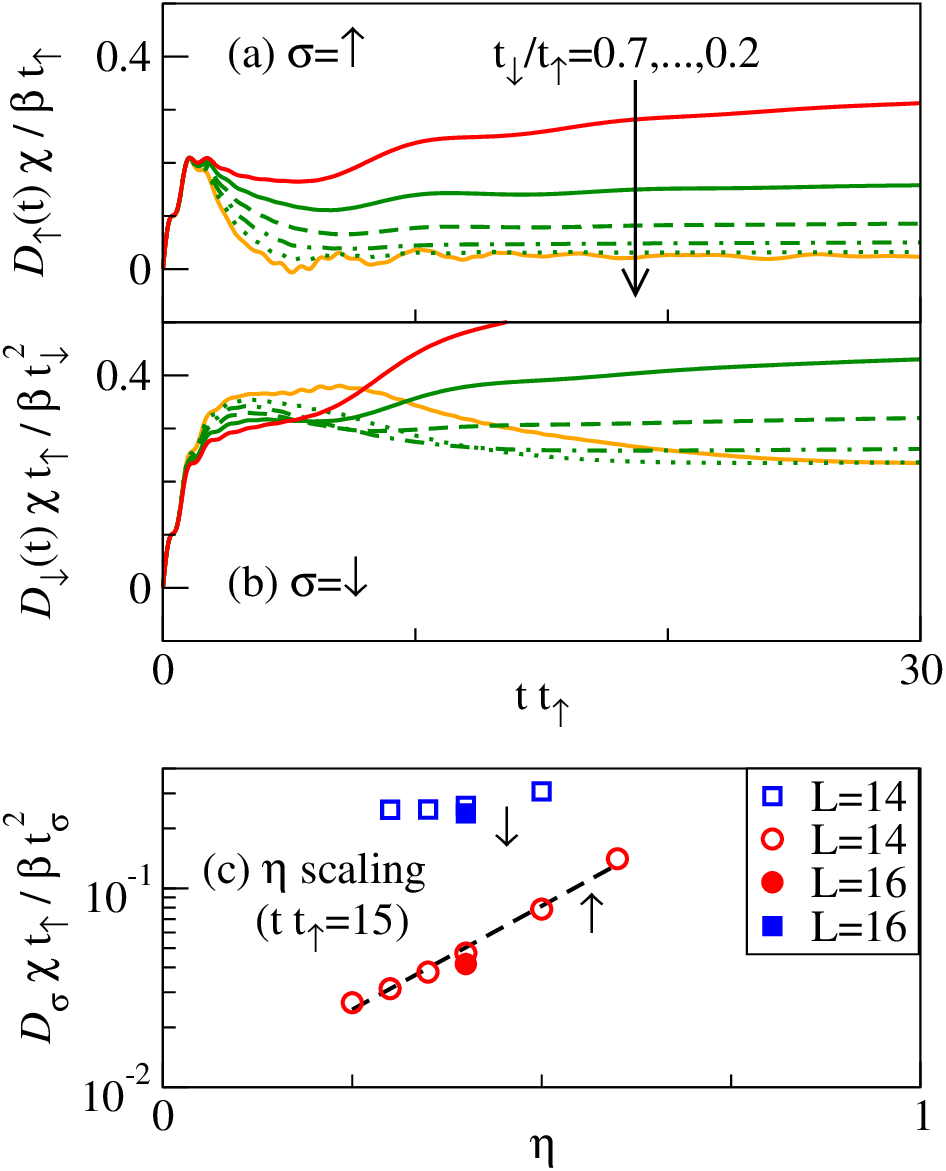}
\end{center} 
\caption{(Color online) Time dependence of the diffusion constant
${\cal D}_\sigma(t)$ for (a) $\sigma=\uparrow$, (b) $\sigma=\downarrow$ for
various $\eta = t_\downarrow/ t_\uparrow = 0.7, \ldots, 0.2$, a single
$U/t_\uparrow = 8$, fixed $L=14$, and high temperature $\beta \, t_\uparrow
\to 0$. Apparently, ${\cal D}_\uparrow(t)$ is very sensitive to varying $\eta$,
in contrast to $D_\downarrow(t)$. For imbalance $\eta \leq 0.6$, a plateau of
${\cal D}_\sigma(t)$ can be already seen for the $L$ depicted. (c) $\eta$
scaling of the plateau value for both components, as extracted at the point
$t \, t_\uparrow = 15$, in a semi-log plot. As a guide to the eyes, an
exponential (dashed line) is indicated. Note that ${\cal D}_\sigma \chi =
\sigma_{\sigma,\text{dc}}$.} 
\label{Fig11}
\end{figure}

We eventually discuss the scaling of transport coefficients as a function of
imbalance $\eta = t_\downarrow/t_\uparrow$. For the $\eta$ discussed below,
extracting the dc conductivity $\sigma_{\sigma,\text{dc}}$ as $\text{Re} \,
\sigma_\sigma(\omega \to 0)$ for finite $L$ is equivalent to determining the
plateau value of the time-dependent diffusion constant ${\cal D}_\sigma(t)$.
Therefore, we focus on an analysis of ${\cal D}_\sigma(t)$, which can be concisely
summarized for various $\eta$.

In Fig.~\ref{Fig11}(a) we show the time-dependent diffusion constant 
${\cal D}_\uparrow(t)$ of the light component for different $\eta = 0.7, \ldots,
0.2$, a single $U/t_\uparrow =8$, and fixed system size $L=14$. In
Fig.~\ref{Fig11}(b) we show ${\cal D}_\downarrow(t)$ of the heavy component for the
same set of parameters. Several comments are in order. First, for both $\sigma
= \uparrow, \downarrow$, a plateau of ${\cal D}_\sigma(t)$ is clearly visible
at times $t \, t_\uparrow \sim 15$ for imbalances $0.3 \leq \eta \leq 0.6$. We
have checked that the plateau values ${\cal D}_\sigma \, \chi$ coincide with the
dc conductivity $\sigma_{\sigma,\text{dc}}$, cf.\ Fig.~\ref{Fig10} for $\eta =
0.4$, even though not shown explicitly for all $\eta$. Second, for $\eta > 0.6$,
${\cal D}_\sigma(t) \propto D_\sigma \, t$ due to strong finite-size Drude
weights $D_\sigma$ in the vicinity of the integrable point $\eta = 1$, cf.\
Fig.~\ref{Fig6}. These finite-size effects prevent us from determining the
diffusion constant in the thermodynamic limit $L \to \infty$. Third, for $\eta
< 0.3$, ${\cal D}_\uparrow(t)$ of the light component develops the small oscillations
around zero discussed in the context of the Falicov-Kimball limit $\eta = 0$.
These oscillations prevent us from determining the diffusion constant with
sufficiently high accuracy. Fourth, ${\cal D}_\uparrow(t)$ is much more sensitive
to varying $\eta$ than ${\cal D}_\downarrow(t)$. Note, however, that we depict
${\cal D}_\downarrow(t) / t_\downarrow^2$ rather than ${\cal D}_\downarrow(t)$.
In this way, we do not show the trivial scaling ${\cal D}_\downarrow(t) \propto
t_\downarrow^2$ resulting from the static scaling of the current operator
$j_\downarrow \propto t_\downarrow$.

In Fig.~\ref{Fig11}(c) we depict the $\eta$ dependence  of the plateau values
${\cal D}_\sigma$, visible for $L=14$, in a semi-log plot. While we find
${\cal D}_\downarrow/ t_\downarrow^2 \approx \text{const.}$, we observe a
decrease of ${\cal D}_\uparrow$ as $\eta$ decreases, consistent with a simple
exponential function. If we {\it assume} that this scaling continues to small
$\eta$ beyond the $\eta$ range accessible, this assumption would imply the
absence of a diffusion-localization transition at $\eta \neq 0$, consistent
with the conclusions of \cite{yao15}. However, based on our results in
Fig.~\ref{Fig11}(c), we cannot exclude the onset of many-body localization
and a sudden drop of ${\cal D}_\uparrow$ to zero at finite but small $\eta$,
as suggested in previous works \cite{grover14,schiulaz14}. Nevertheless,  we can
constrain the existence of a possibly localized regime to $\eta \ll 0.25$.

\section{Summary and Outlook}
\label{sec:sum}
In this work we studied finite-temperature charge transport in the 
one-dimensional repulsive Hubbard model at half filling. Using the method of
dynamical quantum typicality, we were able to access system sizes much
larger than what can be reached with full exact diagonalization, and with
no restriction on the accessible time scales. This allowed us to extract
the finite-size dependent Drude weight from the time dependence of current
autocorrelation functions. The analysis of the finite-size dependencies
indicated a vanishing Drude weight in the thermodynamic limit, in agreement
with \cite{carmelo12}. We further computed the optical conductivity
and provided evidence that it is (i) a smooth function of $\omega$ at low
frequencies and in the thermodynamic limit and (ii) that the dc conductivity
is indeed finite, the latter in agreement with \cite{karrasch14a}.

As an example of a non-integrable model, we considered the mass-imbalanced
Hubbard chain. This model has recently been discussed in the context of
many-body localization in translationally invariant systems \cite{grover14,
schiulaz14,yao15}. We demonstrated the absence of a Drude weight for large
$L$, as expected for a non-integrable system. Our results for inverse mass
ratios of $\eta \gtrsim 0.25$ indicated a small dc conductivity, that appears
to vanish exponentially fast as a function of decreasing $\eta$. At
intermediate $\eta$, the system is thus a normal diffusive conductor, while
at small $\eta$, the emergence of small long-time oscillations
in the current autocorrelation function give rise to slightly anomalous
transport, in line with the conclusions of Ref.~\cite{yao15}.

Extensions of our work comprise the study of finite-temperature charge and
spin transport in one-dimensional strongly correlated electron systems. For
instance, there is an intriguing prediction on the role of spin drag in one
dimension, which has been claimed to give rise to diffusive spin transport,
while charge transports remains ballistic at finite temperature \cite{polini07}.
Such questions as well as other effects due to a coupling of the various
transport channels in the Hubbard model and its variants constitute a rich
playground for future work. 

{\it Acknowledgment.} We thank C. Karrasch for sending us tDMRG data
and very helpful comments. We gratefully acknowledge the computing time
granted by the JARA-HPC Vergabegremium and provided on the JARA-HPC Partition
part of the supercomputer JUQUEEN at Forschungszentrum J\"ulich. R.S.\ thanks
the Arnold-Sommerfeld-Center for Theoretical Physics, LMU Munich, for its kind
hospitality. This work was also supported in part by National Science Foundation
Grant No. PHYS-1066293 and the hospitality of the Aspen Center for Physics.


%

\end{document}